# ChatGPT: Vision and Challenges

Sukhpal Singh Gill[1] and Rupinder Kaur[2]

[1]School of Electronic Engineering and Computer Science, Queen Mary University of London, UK
[2]Department of Science, Kings Education, London, UK
*s.s.gill@qmul.ac.uk, rupinderchem@gmail.com*

**Abstract:** Artificial intelligence (AI) and machine learning have changed the nature of scientific inquiry in recent years. Of these, the development of virtual assistants has accelerated greatly in the past few years, with ChatGPT becoming a prominent AI language model. In this study, we examine the foundations, vision, research challenges of ChatGPT. This article investigates into the background and development of the technology behind it, as well as its popular applications. Moreover, we discuss the advantages of bringing everything together through ChatGPT and Internet of Things (IoT). Further, we speculate on the future of ChatGPT by considering various possibilities for study and development, such as energy-efficiency, cybersecurity, enhancing its applicability to additional technologies (Robotics and Computer Vision), strengthening human-AI communications, and bridging the technological gap. Finally, we discuss the important ethics and current trends of ChatGPT.

**Keywords:** ChatGPT, Artificial Intelligence, Machine Learning, Chatbot, GPT, Generative AI

## 1. Introduction

Over the past few years, language models have benefited greatly from the rapid development of Artificial Intelligence (AI) and Natural Language Processing (NLP), making them more accurate, flexible, and useful than ever before [1]. The term "Generative AI" is used to describe a subset of AI models that can generate new information by discovering relevant trends and patterns in already collected information. These models may produce work in a wide range of media, from written to visual to audio [2]. To analyse, comprehend, and produce material that accurately imitates human-generated outcomes, Generative AI models depend on deep learning approaches and neural networks. OpenAI's ChatGPT is one such AI model that has quickly become a popular and versatile resource for a number of different industries. Its humanoid text generation is made possible by its foundation in the Generative Pre-trained Transformer (GPT) architecture [3]. It has the ability to comprehend and produce a broad variety of words since it has been training on an extensive amount of text data. Linguistic transformation, summarised text, and conversation production are just some of the applications that can benefit from its capacity to create natural-sounding content. ChatGPT can be trained to do a variety of activities, including language recognition, question answering, and paragraph completion. It's also useful for building chatbots and other conversational interfaces. In a nutshell, ChatGPT is a robust NLP model that can comprehend and create natural language for a wide range of applications, including text production, language understanding, and interactive programmes [4].

It is essential to ChatGPT's function in promoting scientific research to comprehend its genesis and evolution. To clarify, the ChatGPT is not a Generative Adversarial Network (GAN) model but rather a linguistic model built on the GPT architecture, which is relevant here. GPT models are tailored to NLP activities including text production and language comprehension, as opposed to GANs, which are more commonly employed for activities including picture generation [5]. The origins of ChatGPT lie in NLP, a subfield of AI that aims to teach computers to comprehend and produce human speech. The motivation behind developing ChatGPT was to establish a powerful and flexible AI language model that could help with a wide range of activities, such as text production, translation, and data analysis.

### 1.1 Motivation and Our Contributions

In order to address some of the limitations of prior sequence-to-sequence models for NLP, ChatGPT was built on the foundation of the Transformer architecture. This innovative design made it possible to make powerful language models like OpenAI's GPT series, which included GPT-2 and GPT-3, which were the versions that came before ChatGPT [6]. The GPT-3.5 architecture is the basis for ChatGPT; it is an improved version of OpenAI's GPT-3 model. Even though GPT-3.5 has fewer variables, nevertheless produces excellent results in many areas of NLP, such as language understanding, text generation, and machine translation [6]. ChatGPT was trained on a massive body of text data and fine-tuned on the goal of creating conversational replies, allowing it to create responses to user inquiries that are strangely similar to those of a person.

The main objectives of this work are:
- To present the roadmap and outlook of ChatGPT.
- To investigate the capabilities of ChatGPT for strengthening human-AI communications.
- To discuss the notable functions of ChatGPT, its popular applications and ethics.
- To examine the advantages of bringing everything together through ChatGPT and IoT.
- To highlight the current trends & research challenges of ChatGPT.

### 1.2 Article Structure

The rest of the paper is structured as follows: *Section 2* presents the background of ChatGPT. *Section 3* discusses the notable functions of ChatGPT. *Section 4* gives the roadmap for the applications of ChatGPT. *Section 5* discusses the advantages of bringing everything together through ChatGPT and Internet of Things (IoT). *Section 6* highlights the future directions and research opportunities. Further, *Section 7* presents the current trends and *Section 8* discusses the important ethics. Finally, Section 9 concludes the article.

## 2. Background and Foundations

The revolutionary models GPT-2, GPT-3, and ultimately ChatGPT were developed by OpenAI, which has been at the cutting edge of AI innovation [51, 52, 53].

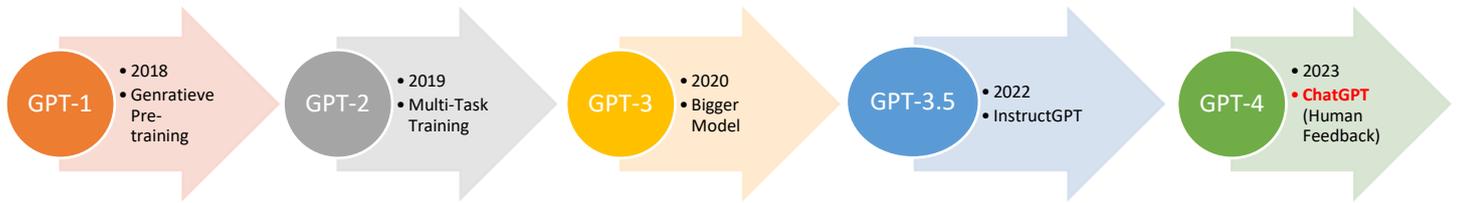

Figure 1: The Timeline of GPTs from the first version (GPT-1) to the latest version (ChatGPT)

OpenAI maintained its study and development efforts after the success of GPT-3, eventually resulting in ChatGPT, which is based on the GPT-4 architecture [7]. ChatGPT is optimised for conversational tasks; it outperforms GPT-3 in terms of contextual comprehension, answer creation, and consistency [51]. OpenAI extended its research and development activities after the release of GPT-3, eventually resulting in ChatGPT, which is based on the GPT-4 model [52]. ChatGPT is optimised for conversational activities; it outperforms GPT-3 in terms of contextual comprehension, answer creation, and coherence [2]. GPT models are developed to create coherent and human-like natural language text, including phrases, paragraphs, and complete papers [53]. In order to perform well on subsequent assignments like text categorization and question answering, GPT models must first be pre-trained on massive volumes of text data. In unsupervised pre-training, the model is trained on a huge body of text data, including that found in textbooks or online, without the use of tags or comments. The GPT model is taught to predict the next word in a text sequence by examining the words that came before it in the training data. In the field of NLP, this is known as a language modelling job [3]. The model trains to recognise and generalise linguistic trends, including syntax, vocabulary, and logic, by training on a vast body of text data. If the GPT model is given a smaller labelled dataset after pre-training, it may be modified on a single downstream job by updating its weights and biases to better match that task [4]. If categorization of text is required to be performed downstream, for instance, the model may be educated to determine which label best fits a particular piece of input text. Figure 1 shows the timeline of GPTs from the first version (GPT-1) to the latest version (ChatGPT).

**2.1) GPT-1**: It is the initial version of the GPT programming language, and it was made available to the public in 2018 [7]. The foundation of the system was the Transformer Neural Network architecture, which was developed specifically for use in NLP applications such as language modelling and machine translation. Using a language modelling job, GPT-1 was first pre-trained on a vast collection of text data that includes documents, papers, and online content. The model was taught to anticipate the next word in a string of text based on the words that came before it. The enormous collection of text data served as a training ground for GPT-1 to acquire word patterns and their interconnections [8]. GPT-1 might be optimised to perform particular downstream tasks including language translation, emotion analysis, and text categorization after initial training. In order to train the model to accurately predict the sentiment of text input, for instance, it would need access to a labelled dataset of text data. When contrasted with subsequent iterations of the GPT model, GPT-1's somewhat limited parameter set of 117 million seems quite modest [8]. Notwithstanding its modest size, GPT-1 proved the efficacy of pre-training on huge volumes of text data for increasing comprehension of languages by achieving outstanding outcomes on a broad spectrum of tasks requiring NLP.

**2.2) GPT-2:** It was a substantial improvement over GPT-1 as among the greatest language models available at the point of its release. It featured 1.5 billion parameters [9]. Pre-trained with a language modelling job, GPT-2 was fed data from a large corpus of textual resources such as text books, articles, and online content. Just like GPT-1, this model was programmed to guess what comes next in a text by looking at what came before. GPT-2, on the other hand, generated longer and more cohesive text sequences and showed more generalizability across tasks and contexts. GPT-2 might be refined for a number of downstream tasks after initial training, including text categorization, emotion analysis, and question answering [10].

**2.3) GPT-3**: It is more significant and efficient than its predecessor, GPT-2 with 175 billion parameters [10]. Using a language modelling job, GPT-3 was trained on a large collection of text data that comprised books, articles, and online content. The model was taught to anticipate the next word in a string of text based on the words that came before it, and it has since been used to produce convincing, natural-sounding prose. GPT-3's versatility lies in its capacity to handle several NLP tasks while requiring job-specific training data. These tasks include text categorization, sentiment analysis, and question answering. It happens because the model may pick up a variety of language elements and patterns from its pre-training data, allowing it to generalise to a broad variety of activities and contexts. In addition, GPT-3 has a number of cutting-edge features, including multi-task learning and few-shot learning, which permit the system to pick up and master new jobs with minimal training data [11]. Due to its many useful features, GPT-3 is a language model for NLP that can be applied to many different domains. Chatbots, language translation, content production, as well as code generation are just some of the numerous practical applications that have made use of GPT-3.

Employing a language modelling job, ChatGPT has already been pre-trained on a huge collection of text data [1]. This data includes books, papers, and online content. ChatGPT is able to effectively generate logical and authentic replies in a discussion since it is pre-trained to grasp the patterns and links among phrases and words in natural language [2]. Figure 2 shows a general architecture of ChatGPT, when a client interacts with ChatGPT to ask a query, the platform is able to deduce the user's purpose using Natural Language Understanding (NLU). The appropriate data is fetched from the underpinning Knowledge Base. Natural Language Generation (NLG) then sums up a response to the client based on this generated response. The discussion history is saved so that future interactions may be responded to and tailored to the individual. In order to enhance the future quality of answers, Reinforcement Learning is used to collect customer feedback and take appropriate actions.

**2.4) GPT-4:** It was released by OpenAI recently, which includes substantial advances in deep learning scalability [7]. This new model is a comprehensive multidimensional linguistic framework that takes in images and text and produces written outcomes. GPT-4 has exhibited human-level performance on a variety of professional and academic standards, even if it might not be as proficient as humans in everyday situations [13]. For example, compared to GPT-3.5, it has attained a score in the highest 10% of participants in tests on a virtual legal examination [12]. There remains an opportunity for enhancements to GPT-4's factuality, steerability, and remaining within the provided restrictions, but after six months of incremental alignment using lessons from OpenAI's adversarial evaluation programme and ChatGPT, the model achieved its best-ever efficiency [57].

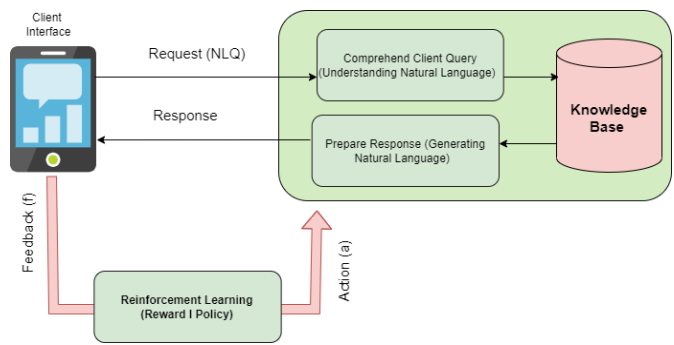

Figure 2: General Architecture of ChatGPT

Text generation, question answering, language translation, and emotion evaluation are merely some of the many NLP tasks in which GPT models have proven themselves to be cutting-edge performers. Chatbots, aiding customers, and material production are just some of the practical applications identified in the literature [1] [2] [5].

## 3. Functions of ChatGPT: Beyond Perspective

Because of its sophisticated and flexible design, ChatGPT can be used for a wide variety of NLP tasks. Some of the ways in which it is able to revolutionise how humans interact with machines are as follows [2, 4, 5, 6, 10, 11, 13, 58]: context-specific awareness; linguistic generating skills; task flexibility; multinational competence; scalability; non existent-shot and several-shot training; and enhancing capability. The most notable functions of ChatGPT are explained below:

*3.1 Cognitive Comprehension:* ChatGPT's capability to grasp conversational contexts is one of its greatest strengths. ChatGPT's authentic and intriguing conversations with customers are a result of its ability to understand the significance of statements and words.

*3.2 Talent to Generate Languages:* ChatGPT's language-generating skills are top-notch; the generated content is logical, appropriate, and free of mistakes in grammar. It may be utilised for tasks like writing materials, summarising, and revising because of its proficiency with creating text.

*3.3 Power to Change Tasks:* ChatGPT's flexibility throughout sectors and disciplines makes it a useful tool for a broad variety of applications. It may be fine-tuned to serve a variety of purposes, including but not limited to client service, document generation, coaching, interpreting, and beyond. Because of its flexibility, ChatGPT may be used by programmers to build specialised applications.

*3.4 Competence in Multiple Languages:* ChatGPT's linguistic flexibility makes it suitable for international deployment and widens its potential user pool. Its ability to translate text, analyse user emotions, and produce material in several languages are all made possible by this feature.

*3.5 The Ability to Grow:* ChatGPT can be scalable in terms of both processing resources and turnaround times because of its modular design. Because of its capacity for growth, it may be used for tasks of varied sizes, from personal endeavours to enterprise-wide programmes.

*3.6 Short-Chance and No-Chance Training:* In order to grasp unfamiliar duties with no lengthy training, ChatGPT is capable of nil-shot and quick learning. The framework may acquire novel tasks with just a few scenarios in few-shot acquiring knowledge, while in zero-shot teaching it can create replies for tasks it hadn't encountered previously. This capability shortens the creation cycle by reducing the requirement for big labelled datasets and considerable refinement.

*3.7 Tweaking*: ChatGPT's tweaking capabilities are essential, enabling programmers to tailor the model to a variety of applications. ChatGPT's replies are more precise and pertinent since the model was trained using a smaller data set that was specifically designed for the intended use case. The ability to fine-tune ChatGPT allows programmers to build extremely specialised recommendations.

*3.8 Rapid Development of ChatGPT:* Optimising the consumer's engagement and facilitating productive interaction when working with AI models like ChatGPT depends heavily on quick implementation. Consumers may direct the AI model to produce more precise, pertinent and helpful replies by quick programming.

## 4. Applications: Roadmap and Outlook

Due to its adaptability and superior NLP skills, ChatGPT has found use outside of the academic research community. In this section, we discuss how ChatGPT can be used in a variety of applications (as shown in Figure 3) illustrating the ways in which the platform can revolutionise business processes, bolster collaboration, and spark new ideas.

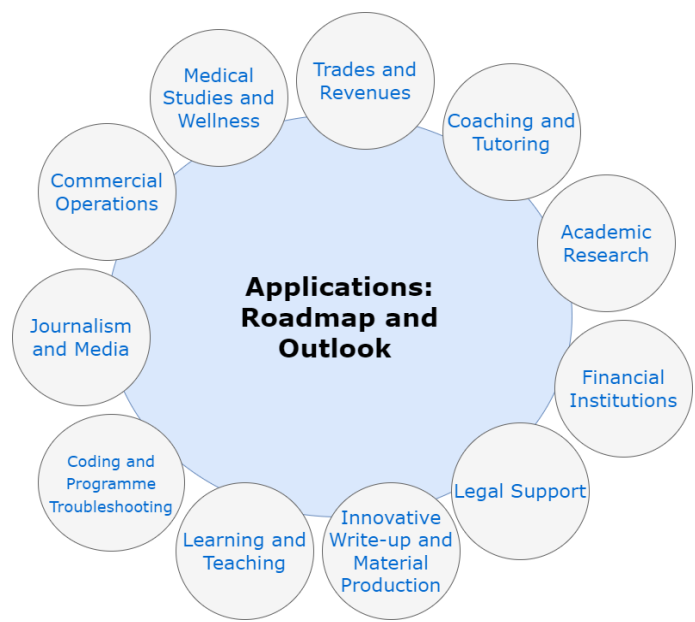

Figure 3: Applications of ChatGPT

### 4.1 Medical Studies and Wellness

ChatGPT's utility in medicine and healthcare [14] lies in its ability to (i) aid in diagnosis by analysing patient records, health status, and indications to produce treatment strategies tailored to each patient's unique desires and requirements, (ii) summarise and synthesise clinical studies to support a research-based practices, (iii) offer medical knowledge and guidance to patients in a simple and digestible manner, and (iv) encourage interaction among medical experts. Further, ChatGPT can be utilised to create robots that can aid with patient evaluation, which is the process by which medical professionals assess the severity of a patient's illness and decide what treatment is necessary [15]. Moreover, ChatGPT can be leveraged to create tools that help doctors diagnose patients and provide care [16]. ChatGPT is able to aid doctors in diagnosing and treating patients by analysing individual data and concerns [35]. Finally, ChatGPT can be utilised to create tools that improve healthcare education.

### 4.2 Trades and Revenues

ChatGPT's applicability in the world of business and finances can be broken down into several categories [17, 18]: (i) the automated preparation of accounting documents and market assessment snippets; (ii) the performance of emotion analysis on consumer ratings and comments to guide new product development and advertising tactics; (iii) the automatic production of personalised recommendation for investments according to user risk histories and monetary targets; (iv) the aiding of the formulation of enterprise plans, advertising collateral, and other created material; and (v) the age of personalised expenditure suggestions. ChatGPT can be utilised to create bots for usage in customer support that can answer questions, offer suggestions, and even complete purchases. A great deal of financial information may be analysed with ChatGPT to reveal trends and patterns as well as views on current markets and movements. ChatGPT's ability to analyse economic data and give recommendations for investments is a boon to companies and financiers. Using ChatGPT, one can create anti-fraud and anti-money-laundering systems. ChatGPT might assist banks and other banking institutions save money by analysing transaction data for trends which could suggest illegal behaviour. ChatGPT helps organisations and the financial sector by analysing financial data and delivering perspectives on their financial health.

### 4.3 Legal Support

ChatGPT can support lawyers in creating legally binding paperwork such as agreements, pleas, and affidavits [19]. ChatGPT can be used to summarise and synthesise legal texts like agreements, laws, and court judgements. ChatGPT can analyse and anticipate the outcomes of legal conflicts based on previous information and judicial precedents, and deliver rapid and precise responses to legal inquiries based on applicable legislation and case law [20]. The following are some examples of how ChatGPT might be used in the legal industry. Researchers in the legal field can benefit from ChatGPT's ability to analyse massive volumes of data, such as legislation, rules, and legal precedent. If an agreement has any inconsistencies or contradictions that might lead to legal trouble, ChatGPT can help you find them. Using ChatGPT, you can create a chatbot that provides legal advice and answers common legal concerns from customers. Legal papers, including briefings, agreements, and other paperwork, may all benefit from the usage of ChatGPT as a writing tool [19]. ChatGPT may assist attorneys in drafting excellent and correct papers by analysing legal facts and making advice.

### 4.4 Innovative Write-up and Material Production

ChatGPT can be used in the fields of writing, editing, and write-up production [21, 22] to (i) produce novel concepts for stories, organise highlights, and personality depictions; (ii) help authors overcome author's block by indicating innovative instructions as well as writing initiates; (iii) produce automatic data for websites, papers, and Facebook and Twitter posts based on particular data parameters and style personal preferences; and (iv) review and revise text to enhance vocabulary, brevity, and readability. Blog entries, social networking material, and advertising text are just a few examples of how ChatGPT may be put to work in the content development process. ChatGPT can produce useful and interesting replies in natural language by analysing data on the subject matter, voice, and attitude [21]. Authors that are stuck for inspiration might utilise ChatGPT to create innovative writing suggestions. ChatGPT generates original and thought-provoking suggestions for writing for authors by analysing data on categories, concepts, and story structures. ChatGPT is able to offer author's specific advice on how to improve their writing by analysing data on prevalent genres and story patterns [22].

### 4.5 Learning and Teaching

ChatGPT can be used in the higher education and training [23, 24] business to (i) establish individualised instructional resources and plans of study based on each learner's distinctive requirements and interests, (ii) guide learners through their education in a timely manner, (iii) produce interesting learning materials like tests, engaging tasks, and interactive videos, (iv) support teachers in evaluating student work and giving constructive criticism, and (v) host online discussions and collaborations between students and instructors. ChatGPT may be utilised to give students individualised education plans by analysing information about their choices, abilities and areas for improvement in the classroom [25]. ChatGPT can help children succeed in school by giving individualised suggestions for study resources and activities. For example, a teacher can utilise ChatGPT to get advice on how to improve lessons, methods of instruction, and administration of the learning environment. ChatGPT provides personalised suggestions to instructors based on an analysis of data on educational best practices and student learning results. If you're taking up a language that is unfamiliar, you can put ChatGPT to good use by getting customised suggestions on how to improve your words, phrases, and pronunciation. ChatGPT may assist learners get ready for examinations more efficiently by analysing data on previous achievements and how they like to learn. By analysing data on an individual's educational requirements and making individualised suggestions for private lessons, ChatGPT may be utilised to deliver virtual tutoring services.

### 4.6 Coding and Programme Troubleshooting

ChatGPT may take customer input and output programme excerpts [26]. ChatGPT analyses information about the programming languages, the functionality, and the specifications in order to supply clients with code excerpts that may be utilised to build the desired characteristic or functionality. By analysing information about the language used for programming, computer programmes, and data structures, ChatGPT can help optimise programmes [27]. ChatGPT is a tool that aids programmers by analysing their code for inefficiency and suggesting solutions. By examining information about the language used for programming, the source code organisation, and the resulting error notifications, ChatGPT may be utilised to aid in the debugging process. ChatGPT aids programmers in finding and fixing errors in programming by suggesting tactics and approaches for troubleshooting [28]. ChatGPT may be utilised to help describe code by analysing information about the programming language, the organisation of the code, and the necessary functions. In order to aid with code inspection, ChatGPT may examine information about the programming language, coding norms, and optimal practices. ChatGPT may aid programmers in enhancing the reliability and dependability of their programmes by pointing out possible errors and suggesting solutions.

### 4.7 Journalism and Media

ChatGPT may be employed to help write screenplays, come up with plot ideas, and even come up with lines of dialogue for TV

episodes, films, and playing video games [29]. ChatGPT can provide interesting and relevant replies in natural language by analysing information about the content's type, voice, and presentation. Users may connect with bots, networking sites, and other live interactions with the help of ChatGPT [30]. ChatGPT can give customised solutions that increase interaction and retention by analysing data on user demographics and behaviour. Movies, TV series, and audio may all be suggested to users depending on their tastes with the help of ChatGPT. When utilised for voice acting, ChatGPT may offer advice and inspiration for different voices, dialects, and inflections to employ for different characters. Voice actors can benefit from ChatGPT's personalised advice since the platform analyses data on the character's personality and history to offer specific guidance.

### 4.8 Commercial Operations

By analysing information about customers' actions and desires, ChatGPT can help with lead creation. ChatGPT may increase a company's lead generation and retention of clients through the use of tailored product and service suggestions [18]. With ChatGPT, you may create chatbots to answer customer enquiries, recommend products, and process purchases [31]. ChatGPT delivers personalised suggestions that improve the customer expertise as a whole by analysing data on customer behaviour and preferences. An enormous amount of advertising information may be analysed with ChatGPT, revealing behaviours and patterns that might inform subsequent advertising strategies and initiatives [6]. ChatGPT analyses information about the audience, message, and tone to provide authentic, interesting replies in natural language. As an advertising training tool, ChatGPT may be leveraged to provide salespeople advice on how to best position products, deal with customer objections, and close deals [32].

### 4.9 Financial Institutions

ChatGPT may be applied to create chatbots for usage in customer support that can answer questions, provide suggestions, and even complete purchases [6]. ChatGPT's ability to offer customised suggestions based on consumer needs and past actions is made possible by the information that it collects and analyses. Systems that can spot money laundering and fraud may be developed with the help of ChatGPT [33]. ChatGPT helps banks and other financial institutions avoid losses by analysing transaction data for patterns that may indicate fraudulent behaviour. Investing in management system development can benefit from the use of ChatGPT. ChatGPT's ability to analyse financial data and provide advice helps financial firms make smart investments [34]. Consumers that need help with spending, saving, or paying down debt might benefit from software built using ChatGPT. ChatGPT aids financial institutions in devising strategies to mitigate risks by analysing financial data and identifying potential threats.

### 4.10 Academic Research

The capacity to handle and analyse huge amounts of data is crucial to research in science [32, 34]. ChatGPT has proven essential in revolutionising the way scientists collaborate and make sense of data. Uses of ChatGPT in handling and analysing data are discussed in this section. Examples of such applications include (i) using NLP for obtaining data from scientific publications, (ii) synthesising and summarising massive amounts of information, (iii) automatically identifying emerging trends and patterns in statistics, and (iv) automated modelling and prediction. Science cannot progress without the capacity for processing and analysing massive amounts of data. ChatGPT has proven to have a substantial effect on boosting researchers' interactions with and interpretations of data, hence increasing productivity and revealing previously unknown discoveries. This section delves into ChatGPT's potential game-changing uses in the data processing and interpretation industry. ChatGPT is mostly used for data analysis applications, such as gathering data from scientific research papers [6]. ChatGPT quickly identifies and extracts crucial data points, results, and implications from dissertations using natural language processing algorithms. This capacity allows scientists to efficiently obtain and synthesise data from many sources, speeding up the study process and decreasing the need for labour-intensive traditional literature assessments. ChatGPT's ability to synthesise knowledge from several data sources and provide concise overviews is another way it aids academics in making sense of complicated datasets [32]. ChatGPT helps researchers get an in-depth comprehension of their findings by highlighting recurring themes, trends, and interrelationships throughout the data [34].

### 4.11 Coaching and Tutoring

ChatGPT has an opportunity to transform the way that students acquire scientific knowledge by providing them with individualised instruction. ChatGPT can better help its users grasp and remember concepts by taking into account their unique learning preferences, advantages, and shortcomings [36]. This individualised method of instruction has the potential to eliminate knowledge gaps and provide learners the tools they need to become successful in the Science, Technology, Engineering, and Mathematics (STEM) fields [37]. ChatGPT's impact is not limited to the academic world; it has also helped in popularising science and enhancing classroom instruction. The ability of ChatGPT to revolutionise science learning has been explored, along with the numerous ways it has been used to spread scientific literacy. The potential of ChatGPT's uses in public awareness and education about science to increase the public's scientific knowledge, engagement, and interest is enormous.

## 5. ChatGPT and IoT: Intertwined in the NLP

IoT is changing how the world is engaging with the natural environment in fundamental ways. The potential applications range widely, from smart houses to interconnected vehicles [49]. What happens, then, when we integrate ChatGPT in IoT devices, which is among the most sophisticated models for interpreting natural language? This could be an extraordinary connection, with gadgets now able to talk to one another and understand spoken language. In this section, we discuss the convergence of ChatGPT and the IoT opens up fascinating new avenues for use of technology [38]. The upcoming era of the IoT is more vibrant than ever, with possibilities ranging from voice-controlled home automation to two-way conversations with vehicles. So, there is a need to find its benefits. The phrase "IoT" is commonly used to describe the ever-expanding collection of interconnected electronic gadgets that may exchange data via the web. Connected automobiles, smart appliances, and even factory equipment all fall into this category [39]. Data from these devices may be shared and automated processes can be improved. The goal of the IoT is to provide better connectivity and remote monitoring of devices so that our daily lives are less burdensome and more productive [50]. In simple terms, it's the discipline of adding Internet connectivity to commonplace items so that things can be managed, tracked, and analysed online.

### 5.1 NLP and IoT

NLP is the area of AI that focuses on the study and creation of the English language. IoT devices that use NLP make it easier for humans and machines to communicate in a way that seems more straightforward and simple to both parties [40]. Suppose, for instance, if you could command your smart home gadgets just by speaking to them, rather than using a smartphone app or a physical interface. The use of NLP in these devices makes it feasible for them to comprehend as well as react to conversations between people. This is also true for other IoT gadgets, such as automobiles, factories, and robots. NLP in IoT devices paves the way for higher-level interactions including natural language conversation and question answering. The incorporation of NLP into IoT allows for more organic and straightforward communication between humans and their gadgets [41]. In a nutshell, NLP and IoT are two innovations that enhance one another, allowing for more complex and sophisticated engagements by facilitating greater effortless and instinctive communication between machines and people.

### 5.2 Advantages of integrating ChatGPT into IoT devices

ChatGPT enabled IoT devices' improved comprehension and interaction with human speech. The capacity to use voice commands is a major perk of integrating ChatGPT with IoT devices. This simplifies device control by doing away with the requirement for a dedicated app or controller. Humans can tell it to do something like *"turning off the exhaust fan in the kitchen"* and it is going to do so [42]. The capacity to communicate with the gadget in a natural language is another benefit. This implies the gadget would comprehend your inquiries and answer in a humanoid fashion. The engagement may become more interesting and tailored as a result. Furthermore, ChatGPT enables more complex and refined communication with IoT gadgets, such as customization for tasks like question answering and suggestion making [43]. In conclusion, integrating ChatGPT into IoT gadgets allows deeper and more sophisticated interactions, natural language discussion, and removes the requirement for dedicated apps or remote controllers.

### 5.3 Real-world Applications of IoT and ChatGPT

ChatGPT has several applications, including those seen in smart home technology. Through it, humans can use his voice to adjust the volume, temperature, and on/off status of different household devices [43]. ChatGPT is also used in connected vehicles. The driver can use speaking commands to adjust the temperature, play songs, and follow directions. Using natural language, passengers may have conversations with their vehicles about things like directions and weather conditions. In addition to its application in consumer IoT, ChatGPT has found a home in the realm of Industrial IoT, where it enables the monitoring and management of industrial machinery using speech commands or natural language message boards [44]. Overall, the combination of ChatGPT with IoT enables more easy and logical interaction and management of devices through speaking commands and conversational dialogue, which is especially useful for things such as smart home gadgets, linked autos, and manufacturing machinery.

### 5.4 Thinkable Future Advances and Implications

We are discussing potential avenues for developing and incorporating the ChatGPT's NLP paradigm into IoT gadgets. ChatGPT's future growth may come from its incorporation with additional medical technology, such as smartwatches and other wearable devices. These gadgets may monitor a patient's vital statistics, remind them to take their medications, and even have a conversation with their physician or nurse in real time using only the individual's own words [45]. ChatGPT's embedding with smart city networks is another possible use case. Technologies like traffic signals, parking metres, and mass transit networks may fall under this category. It would be much simpler and more practical to communicate with such devices utilising natural language discourse. In the realm of robotics, for instance, ChatGPT might be used to improve the robots' ability to comprehend and react to language spoken by humans in an increasingly anthropomorphic fashion [46]. ChatGPT might also be utilised in the customer service industry to make chatbots and robotic assistants more human-like in how they communicate with clients. ChatGPT and IoT have enormous possibilities for future growth and usage, with a wide range of possible use cases ranging from medical equipment and urban planning to robots and customer support.

### 5.5 ChatGPT and IoT's Influence on Human Life

We have discussed above how ChatGPT and IoT may function alongside to make technology and communication more natural and transparent. Integrating ChatGPT into IoT devices paves the way for voice control and natural language communication, which in turn makes these gadgets more practical for everyday usage [47]. ChatGPT and IoT have a profound effect on daily lives, making them easier and more productive thanks to remote control and monitoring of equipment and enhanced communication between them. It also paves the way for cutting-edge developments in areas like medical care, urban planning, and robots, all of which stand to vastly enhance our quality of life [48]. The interplay between ChatGPT and the IoT is a significant one that is altering the manner in which we engage with technologies and has the promise to substantially enhance our lives in the years to come.

ChatGPT, OpenAI's sophisticated NLP paradigm, and IoT provide interesting possibilities, as explained in this section. ChatGPT's incorporation into IoT gadgets paves the way for greater complexity and sophisticated conversations, as well as more easy and simple interaction, eliminating the requirement for supplemental applications or handheld devices, and allows natural language discourse. Smart household gadgets, networked automobiles, and factory machinery are only a few instances of IoT and ChatGPT in action.

### 6. Research Opportunities and Lights to the Future

Although ChatGPT has been incredibly helpful in improving discoveries in science, it is important to recognise and tackle the research challenges that have been raised by its implementation. Figure 4 shows the summary of research opportunities and possible future directions. This section investigates these challenges and looks ahead to the potential of ChatGPT in the scientific community in the form of future trends [8, 11, 17, 25, 26, 29, 33, 34, 37, 38, 41, 45, 46, 47]. Issues that often arise while employing ChatGPT in the realm of scientific inquiry include the following:

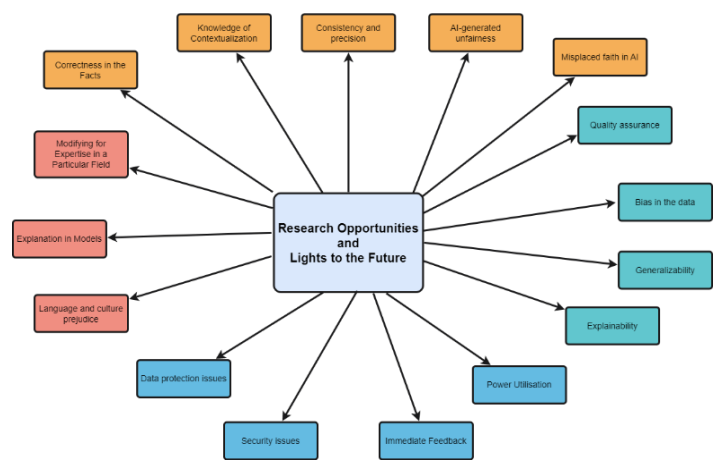

Figure 4: Research Opportunities and Lights to the Future

*6.1 Consistency and Precision:* ChatGPT has demonstrated impressive ability in creating humanoid writing, but it has also been found to make mistakes or provide inaccurate data on rare occasions. Retaining trust in scientific findings depends on the precision and consistency of AI-generated information.

*6.2 AI-generated Unfairness:* The massive volumes of text used to train ChatGPT may have inherent flaws. The AI model may unwittingly spread these biases, which might affect the course of future studies.

*6.3 Misplaced faith in AI:* Dependence excessively on more sophisticated AI models like ChatGPT may lead to a decline in researchers' capacity for autonomous thought and problem solution.

*6.4 Quality Assurance:* ChatGPT can generate outstanding language, but it can also come up with replies that are poor-quality or unsuitable. Continuous surveillance, learning, and improvement are necessary to keep ChatGPT producing excellent content.

*6.5 Bias in the Data:* ChatGPT's results can be affected by the quantity and variety of data that is being trained on it. The adverse impacts in fields such as medical care, law enforcement, and jobs may result from using prejudiced data for training to build predictions.

*6.6 Generalizability:* ChatGPT is frequently inaccurate and has trouble generalising to novel data since it is typically trained on extremely big datasets. The establishment of novel techniques for training is essential for enhancing ChatGPT's generalizability.

*6.7 Explainability:* The ChatGPT model is complicated and hard to comprehend. This might make it tough to deduce the model's decision-making process and spot any inherent flaws.

*6.8 Power Utilisation:* Because of the amount of information they contain, ChatGPT models consume a lot of computational power and may have an adverse effect on the ecosystem. There is much room for improvement in the power effectiveness of ChatGPT models.

*6.9 Immediate Feedback:* Although ChatGPT can produce text in immediate fashion, it may take a while to respond. Many users would benefit from faster and more adaptable ChatGPT.

*6.10 Security Issues:* Intolerance and disinformation are only two examples of the damaging content that ChatGPT might produce. Developing safeguards to stop the creation of such content is crucial.

*6.11 Data Protection Issues:* Users' privacy and security may be at risk since ChatGPT can access so much of their information. The safekeeping and ethical application of user data necessitates the formulation of appropriate policies and laws.

*6.12 Language and Culture Prejudice:* It's possible that ChatGPT has prejudices towards some language and culture combinations, leading to unsuitable or prejudiced replies. More culturally and linguistically inclusive training datasets and assessment measures are needed to combat these prejudices.

*6.13 Explanation in Models:* ChatGPT and other AI language models may offer results that are difficult to interpret and defend. Trust may be sustained and consumers can come to better judgements based on the produced material if these models become more explainable, how they make decisions are more readily apparent, and they provide details about how they operate.

*6.14 Modifying for Expertise in a Particular Field:* ChatGPT has a broad variety of broader skills and understanding, but it may lack the comprehensiveness of specific domain expertise necessary for some tasks. To fully realise their promise, AI language models must be effectively adapted and fine-tuned for specific areas, sectors, and use cases.

*6.15 Knowledge of Contextualization:* While ChatGPT can produce responses that are both logical and conscious of their context, it might find it difficult to comprehend the bigger picture or to remain consistent over the course of conversations that last. Improving the model's capacity to understand and retain meaning over longer chunks of content is a persistent problem that has to be solved.

*6.16 Correctness in the Facts:* ChatGPT and other AI language models may produce unreliable or inaccurate content. Specifically in applications wherein precise facts are vital, like headlines, schooling or medical treatment, it is a key problem to ensure that the produced material is actually valid and compatible with the supplied data.

By resolving these issues and challenges, the AI scientific community will be able to enhance the efficiency, accuracy, and utility of language models including ChatGPT, allowing for the development of further sophisticated and ethical AI-driven applications across a wide range of industries.

## 7. Current Trends

As a growing number of users utilise ChatGPT, it is being exposed to new types of challenges. Students may help advance AI and increase its future usefulness to customers by abandoning traditional schooling in preference for ChatGPT's conveniences. The field of educational technology will also feel ChatGPT's impact. Many educational technology companies are now offering learners a chance to ask queries and have their doubts resolved while also providing them with a subject's fundamentals through ChatGPT. ChatGPT has certain issues, but it will be useful in practise. Therefore, businesses are keen to implement it for financial gain. In the coming future, AI will be able to determine which students will benefit most from working with which instructors, individuals who can not just fill in knowledge gaps but also provide meaningful sources of motivation, mentoring, and advice. The complicated combination of anxiety, hope, worry, amazement, and fright that characterises our society's responses to the quickly expanding AI ecosystem calls for thoughtful consideration in the development of novel AI applications.

### 7.1 Integration with Computer Vision and Robotics

ChatGPT's potential becomes brightest when it's combined with other forms of AI, such computer vision and robotics. Smart and interactive AI systems that combine the linguistic skills of ChatGPT with the perceptual and tangible abilities of computer vision and bots can transform the way humans engage with technologies. Think of a future when you can have a discussion in your native language with a home automation system to adjust the temperature, lighting, and other gadgets, or with a robot that can help you with housework or tasks like grocery shopping. Enhanced natural language creation and a more fluid and intuitive experience for users are two outcomes of the confluence of AI technologies that will make ChatGPT more expert at navigating the nuances of human interaction.

### 7.2 Integration with IoT

ChatGPT and IoT have great potential for expansion and growth, with potential applications spanning from healthcare technology to urban planning to robots and client service. Further, bringing ChatGPT and IoT together is an impressive combination that is altering the way people engage with gadgets and has an opportunity to significantly enhance the way we live in the decades to come.

### 7.3 Cybersecurity Threat

Hackers are leveraging ChatGPT-themed traps to propagate malware throughout Facebook, Instagram, and WhatsApp as demand for generative AI chatbots rises among people across the globe[1]. Meta, Facebook's parent company, released a research on the proliferation of malware disguised as ChatGPT across its various platforms. The company's security researchers have discovered 10 types of malware employing ChatGPT that have attacked customers' devices since March 2023.

### 7.4 Tackling Carbon Footprints of Generative AI

By adjusting the model's framework and advocating for energy-efficient hardware and the usage of renewable energy sources, greenhouse gas emissions might be reduced throughout the development of Generative AI [56]. Furthermore, the amount of inefficient operations may be cut down by optimising the functioning of AI models.

Finally, we speculated on the future of ChatGPT by exploring many prospects for further research and development in terms of sustainability and cybersecurity. These include boosting its applicability to other technologies such as IoT, improving human-AI conversations, and closing the gap between technical advancements.

### 8. A Moral Panic: Ethics

When it comes to the use of AI language models such as ChatGPT, ethics in computer science is a multidimensional field that examines moral and ethical concerns raised by its creation, distribution, and usage [54]. Developing and implementing these technologies in methods which uphold human values and improve the lives of people and communities is crucial. An important ethical issue is how these technologies affect our planet through things like energy usage, electronic trash, and carbon emissions [56]. The environmental effect of computing may be mitigated in a number of ways, including through the creation of energy-effective hardware and software, the encouragement of recycling and proper elimination of electronic trash, and the consideration of product life cycles [59]. As AI systems improve in sophistication, new ethical issues related to AI and machine learning arise. Creating AI systems that uphold human values, behave ethically, and arrive at defensible conclusions is the goal of machine ethics [55]. This encompasses studies of ethical frameworks and rules for AI systems, as well as studies of value synchronisation and explainable AI [60]. Privacy refers to the safeguarding of private information and the guarantee of individual say over its collection, storage, and use [57]. The storage and processing of ever-increasing amounts of personally identifiable information by computer systems has made protecting user privacy an urgent ethical matter.

### 9. Conclusions and Summary

In summary, ChatGPT and huge language models have an upward trajectory and might significantly alter our interactions with technologies. There are many intriguing opportunities for this technology to improve the way we live in significant and helpful ways, from the interaction with other AI technologies to the possibility for enhanced personalisation and customisation to the continuing improvement of language model efficiency. While these innovations have the ability to improve our daily lives, it is our responsibility as humans to evaluate and solve any ethical or societal concerns that may arise as a result of our enthusiastic adoption of them.

**Declaration of competing interest**

The authors declare that they have no known competing financial interests or personal relationships that could have appeared to influence the work reported in this paper.

**References**

[1] Thorp, H. Holden. ChatGPT is fun, but not an author. *Science* 379, no. 6630 (2023): 313-313.
[2] Taecharungroj, Viriya. "What Can ChatGPT Do?" Analyzing Early Reactions to the Innovative AI Chatbot on Twitter. *Big Data and Cognitive Computing* 7, no. 1 (2023): 35.
[3] Biswas, Som S. Role of ChatGPT in public health. *Annals of Biomedical Engineering* (2023): 1-2.
[4] van Dis, Eva AM, Johan Bollen, Willem Zuidema, Robert van Rooij, and Claudi L. Bockting. ChatGPT: five priorities for research. *Nature* 614, no. 7947 (2023): 224-226.
[5] Dash, Bibhu, and Pawankumar Sharma. Are ChatGPT and Deepfake Algorithms Endangering the Cybersecurity Industry? A Review. *International Journal of Engineering and Applied Sciences* 10, no. 1 (2023).
[6] Seghier, Mohamed L. ChatGPT: not all languages are equal. *Nature* 615, no. 7951 (2023): 216-216.
[7] Oxford Analytica. GPT-4 underlines mismatch on AI policy and innovation. *Emerald Expert Briefings oxan-es* (2023).
[8] Welch, Simon. Comparative studies on the human glutamate-pyruvate transaminase phenotypes—GPT 1, GPT 2-1, GPT 2. *Humangenetik* 30 (1975): 237-249.
[9] Saravanan, Shruti, and K. Sudha. GPT-3 Powered System for Content Generation and Transformation. In 2022 Fifth International Conference on Computational Intelligence and Communication Technologies (CCICT), pp. 514-519, 2022.
[10] Zhang, Zhengyan, Xu Han, et al. CPM: A large-scale generative Chinese pre-trained language model. *AI Open* 2 (2021): 93-99.
[11] Dehouche, Nassim. Plagiarism in the age of massive Generative Pre-trained Transformers (GPT-3). *Ethics in Science and Environmental Politics* 21 (2021): 17-23.
[12] Cyphert, Amy B. A human being wrote this law review article: GPT-3 and the practice of law. *UC Davis L. Rev.* 55 (2021): 401.
[13] Dhingra, Sifatkaur, Manmeet Singh et al. Mind meets machine: Unravelling GPT-4's cognitive psychology. *arXiv e-prints* (2023): arXiv-2303.
[14] Sallam, Malik. ChatGPT utility in healthcare education, research, and practice: Systematic review on the promising perspectives and valid concerns. *Healthcare*, vol. 11, no. 6, p. 887, 2023.
[15] Vaishya, Raju, Anoop Misra, and Abhishek Vaish. ChatGPT: Is this version good for healthcare and research?. *Diabetes & Metabolic Syndrome: Clinical Research & Reviews* 17, no. 4 (2023): 102744.
[16] Sallam, Malik. The utility of ChatGPT as an example of large language models in healthcare education, research and practice: Systematic review on the future perspectives and potential limitations. *medRxiv* (2023): 2023-02.
[17] Yue, Thomas, David Au et al. Democratizing financial knowledge with ChatGPT by OpenAI: Unleashing the Power of Technology. *Available at SSRN* 4346152 (2023).
[18] Jain, Varsha, Himanshu Rai, et al. The Prospects and Challenges of ChatGPT on Marketing Research and Practices. *Emmanuel, The Prospects and Challenges of ChatGPT on Marketing Research and Practices* (2023).
[19] Pettinato Oltz, Tammy. ChatGPT, Professor of Law. Professor of Law (February 4, 2023) (2023).
[20] Macey-Dare, Rupert. How ChatGPT and Generative AI Systems will Revolutionize Legal Services and the Legal Profession. *Available at SSRN* (2023).
[21] Hosseini, Mohammad, and Serge PJM Horbach. Fighting reviewer fatigue or amplifying bias? Considerations and recommendations for use of ChatGPT and other Large Language Models in scholarly peer review. (2023).
[22] Kashyap, Ravi, and ChatGPT OpenAI. A First Chat with ChatGPT: The First Step in the Road-Map for AI (Artificial Intelligence) Available at SSRN (2023).

---
[1] https://techcrunch.com/2023/05/03/malware-chatgpt-lures-facebook/

[23] Baidoo-Anu, David, and Leticia Owusu Ansah. Education in the era of generative artificial intelligence (AI): Understanding the potential benefits of ChatGPT in promoting teaching and learning. *Available at SSRN 4337484* (2023).
[24] Kasneci, Enkelejda et al. ChatGPT for good? On opportunities and challenges of large language models for education. *Learning and Individual Differences* 103 (2023): 102274.
[25] Ali, Jamal Kaid Mohammed et al. Impact of ChatGPT on Learning Motivation: Teachers and Students' Voices. *Journal of English Studies in Arabia Felix* 2, no. 1 (2023): 41-49.
[26] Surameery, Nigar M. Shafiq, and Mohammed Y. Shakor. Use ChatGPT to solve programming bugs. *International Journal of Information Technology & Computer Engineering (IJITC)* ISSN: 2455-5290 3, no. 01 (2023): 17-22.
[27] Badruddin Bin Ghazali, Ahmad. Utilising ChatGPT. *BDJ Student, Nature* 30, no. 2 (2023): 5-5.
[28] Fill, Hans-Georg, Peter Fettke, and Julius Köpke. Conceptual Modeling and Large Language Models: Impressions From First Experiments With ChatGPT. *Enterprise Modelling and Information Systems Architectures (EMISAJ)* 18 (2023): 1-15.
[29] Pavlik, John V. Collaborating With ChatGPT: Considering the Implications of Generative Artificial Intelligence for Journalism and Media Education. *Journalism & Mass Communication Educator* (2023): 10776958221149577.
[30] Biswas, Som. Role of chatGPT in Journalism: According to chatGPT. *Available at SSRN* 4405396 (2023).
[31] George, A. Shaji, and AS Hovan George. A review of ChatGPT AI's impact on several business sectors. *Partners Universal International Innovation Journal* 1, no. 1 (2023): 9-23.
[32] Dwivedi, Y. K, et al. (2023). So what if ChatGPT wrote it? Multidisciplinary perspectives on opportunities, challenges and implications of generative conversational AI for research, practice and policy. *International Journal of Information Management*, 71, 102642.
[33] Ali, Hassnian, and Ahmet Faruk Aysan. What will ChatGPT Revolutionize in Financial Industry?. *Available at SSRN* 4403372 (2023).
[34] Yue, Thomas, David Au, Chi Chung Au, and Kwan Yuen Iu. Democratizing financial knowledge with ChatGPT by OpenAI: Unleashing the Power of Technology. *Available at SSRN* 4346152 (2023).
[35] Lin, J.C., Younessi, D.N., Kurapati, S.S. et al. Comparison of GPT-3.5, GPT-4, and human user performance on a practice ophthalmology written examination. Eye, *Nature* (2023).
[36] Castro Nascimento, Cayque Monteiro, and André Silva Pimentel. Do Large Language Models Understand Chemistry? A Conversation with ChatGPT. *Journal of Chemical Information and Modeling* (2023).
[37] Kung, Tiffany H et al. Performance of ChatGPT on USMLE: Potential for AI-assisted medical education using large language models. *PLoS digital health* 2, no. 2 (2023): e0000198.
[38] Haleem, Abid, Mohd Javaid, and Ravi Pratap Singh. An era of ChatGPT as a significant futuristic support tool: A study on features, abilities, and challenges. *BenchCouncil transactions on benchmarks, standards and evaluations* 2, no. 4 (2022): 100089.
[39] Iftikhar, Linta. Docgpt: Impact of chatgpt-3 on health services as a virtual doctor. *EC Paediatrics* 12, no. 1 (2023): 45-55.
[40] Rani, Paul Jasmin, Jason Bakthakumar, B. Praveen Kumaar, U. Praveen Kumaar, and Santhosh Kumar. Voice controlled home automation system using natural language processing (NLP) and internet of things (IoT). In 2017 Third International Conference on Science Technology Engineering & Management, pp. 368-373. IEEE, 2017.
[41] Alexakis, George, Spyros Panagiotakis, Alexander Fragkakis, Evangelos Markakis, and Kostas Vassilakis. Control of smart home operations using natural language processing, voice recognition and IoT technologies in a multi-tier architecture. *Designs* 3, no. 3 (2019): 32.
[42] Aydın, Ömer, and Enis Karaarslan. OpenAI ChatGPT generated literature review: Digital twin in healthcare. *Available at SSRN* 4308687 (2022).
[43] Wang, Fei-Yue, Jing Yang, Xingxia Wang, Juanjuan Li, and Qing-Long Han. Chat with chatgpt on industry 5.0: Learning and decision-making for intelligent industries. *IEEE/CAA Journal of Automatica Sinica* 10, no. 4 (2023): 831-834.
[44] Kasneci, Enkelejda, Kathrin Seßler, Stefan Küchemann, Maria Bannert, Daryna Dementieva, Frank Fischer, Urs Gasser et al. ChatGPT for good? On opportunities and challenges of large language models for education. *Learning and Individual Differences* 103 (2023): 102274.
[45] Deng, Jianyang, and Yijia Lin. The Benefits and Challenges of ChatGPT: An Overview. *Frontiers in Computing and Intelligent Systems* 2, no. 2 (2022): 81-83.
[46] Mijwil, Maad M., Kamal Kant Hiran, Ruchi Doshi, Manish Dadhich, Abdel-Hameed Al-Mistarehi, and Indu Bala. ChatGPT and the Future of Academic Integrity in the Artificial Intelligence Era: A New Frontier. *Al-Salam Journal for Engineering and Technology* 2, no. 2 (2023): 116-127.
[47] Rousseau, Henri-Paul. From Gutenberg to Chat GPT: The Challenge of the Digital University. No. 2023rb-02. CIRANO, 2023.
[48] Gill, Sukhpal Singh, et al. AI for next generation computing: Emerging trends and future directions. *Internet of Things* 19 (2022): 100514.
[49] Biswas, Som S. Role of chat gpt in public health. Annals of Biomedical Engineering (2023): 1-2.
[50] Singh, Raghubir, et al. Edge AI: a survey. Internet of Things and Cyber-Physical Systems (2023).
[51] OpenAI, https://openai.com/, Available Online, Accessed on April, 2023.
[52] OpenAI Blog, https://openai.com/blog/chatgpt, Available Online, Accessed on April, 2023.
[53] ChatGPT, https://chat.openai.com/chat, Available Online, Accessed on April, 2023.
[54] Liebrenz, Michael, Roman Schleifer, Anna Buadze, Dinesh Bhugra, and Alexander Smith. Generating scholarly content with ChatGPT: ethical challenges for medical publishing. *The Lancet Digital Health* 5, no. 3 (2023): e105-e106.
[55] Curtis, Nigel. To ChatGPT or not to ChatGPT? The impact of artificial intelligence on academic publishing. *The Pediatric Infectious Disease Journal* 42, no. 4 (2023): 275.
[56] An, Jiafu, Wenzhi Ding, and Chen Lin. ChatGPT: tackle the growing carbon footprint of generative AI. *Nature* 615, no. 7953 (2023): 586-586.
[57] Sanderson, Katharine. GPT-4 is here: what scientists think. *Nature* 615, no. 7954 (2023): 773-773.
[58] Xie, H. The promising future of cognitive science and artificial intelligence. *Nature Reviews Psychology* 2, 202 (2023)
[59] Editorials, N. Tools such as ChatGPT threaten transparent science; here are our ground rules for their use. *Nature* 613 (2023): 612.
[60] Prunkl, Carina EA et al. Institutionalizing ethics in AI through broader impact requirements. *Nature Machine Intelligence* 3, no. 2 (2021): 104-110.